\documentclass[conference]{IEEEtran}
\IEEEoverridecommandlockouts
\usepackage{cite}
\usepackage{verbatim}
\usepackage{amsmath,amssymb,amsfonts}
\usepackage{algorithmic}
\usepackage{graphicx}
\usepackage{textcomp}
\usepackage{xcolor}
\def\BibTeX{{\rm B\kern-.05em{\sc i\kern-.025em b}\kern-.08em
    T\kern-.1667em\lower.7ex\hbox{E}\kern-.125emX}}
\usepackage{xurl}                   
\usepackage[hidelinks]{hyperref}    
\hypersetup{breaklinks=true}        

\usepackage{url}
\usepackage{hyperref}
\begin{document}

\title{The Orchestration of Multi-Agent Systems: Architectures, Protocols, and Enterprise Adoption}

\author{
\IEEEauthorblockN{Apoorva Adimulam}
\IEEEauthorblockA{
\textit{Applied Agentic AI} \\
\textit{Skan AI}\\
Menlo Park, USA \\
\href{mailto:apoorva.adimulam@skan.ai}{apoorva.adimulam@skan.ai}
}
\and
\IEEEauthorblockN{Rajesh Gupta}
\IEEEauthorblockA{
\textit{Applied Agentic AI} \\
\textit{Skan AI}\\
Menlo Park, USA \\
\href{mailto:rajesh.gupta@skan.ai}{rajesh.gupta@skan.ai}
}
\and
\IEEEauthorblockN{Sumit Kumar}
\IEEEauthorblockA{
\textit{Applied Agentic AI} \\
\textit{Skan AI}\\
Bengaluru, India \\
\href{mailto:sumit.kumar@skan.ai}{sumit.kumar@skan.ai}
}
}

\maketitle
\begin{abstract}
Orchestrated multi-agent systems represent the next stage in the evolution of artificial intelligence, where autonomous agents collaborate through structured coordination and communication to achieve complex, shared objectives. This paper consolidates and formalizes the technical composition of such systems, presenting a unified architectural framework that integrates planning, policy enforcement, state management, and quality operations into a coherent orchestration layer. Another primary contribution of this work is the in-depth technical delineation of two complementary communication protocols—the Model Context Protocol, which standardizes how agents access external tools and contextual data, and the Agent-to-Agent protocol, which governs peer coordination, negotiation, and delegation. Together, these protocols establish an interoperable communication substrate that enables scalable, auditable, and policy-compliant reasoning across distributed agent collectives. Beyond protocol design, the paper details how orchestration logic, governance frameworks, and observability mechanisms collectively sustain system coherence, transparency, and accountability. By synthesizing these elements into a cohesive technical blueprint, this paper provides comprehensive treatments of orchestrated multi-agent systems—bridging conceptual architectures with implementation-ready design principles for enterprise-scale AI ecosystems.

\end{abstract}


\section{Introduction}
The landscape of LLM-powered agents has undergone a marked transformation. Early deployments prioritized isolated, task-specific agents, highly specialized systems with narrow operating scopes. However, contemporary trends point toward ecosystems of collaborating agents. This transition mirrors broader developments in distributed computing, where value emerges less from individual capabilities and more from orchestrated interactions within a collective. 

Several technical drivers explain why the pivot to multi-agent architectures is emerging now including: 
\begin{itemize}
\item scalability limits of LLMs, where context length and reasoning bottlenecks constrain performance
\item the need for specialization versus generalization, enabling modular agents optimized for specific domains to be composed dynamically
\item advances in communication protocols, with message-passing abstractions and nascent standards for inter-agent APIs
\item  economic efficiency, as distributed collectives of smaller agents often outperform costly all-purpose deployments as demonstrated in \cite{halo}, \cite{evolvingorch}
\end{itemize}

Recent industry signals underscore the momentum of this transition. At the enterprise level, PwC has positioned its Agent OS\cite{pwc25} as a switchboard for multi-agent coordination, emphasizing composability and interoperability across enterprise functions. Similarly, Accenture’s Trusted Agent Huddle\cite{accenture25} introduces governance mechanisms for secure, cross-organizational workflows and aligns with the emerging Agent-to-Agent (A2A) protocol. At the same time, research and development within the technical ecosystem has accelerated with frameworks such as LangChain, AutoGen, IBM Watsonx Orchestrate, and Google’s Agent Development Kit providing modular infrastructure for coordination, negotiation, and role-based task delegation. Collectively, these initiatives signal rapid movement toward standardization and operational readiness. 

The remainder of this paper progresses from conceptual foundations to practical realization. Section II traces the evolution of agentic systems toward orchestrated collectives, while Section III establishes the architectural composition of multi-agent systems and outlines their essential components. Sections IV–VII expand on these components in depth, examining how specialized agent roles, orchestration logic, communication protocols, and governance mechanisms together form the technical backbone of orchestrated intelligence. Section VIII presents real-world case studies that demonstrate the practical impact of these architectures across industries. Section IX discusses the challenges and risks associated with scaling multi-agent systems, and the future research directions aimed at improving efficiency, reliability, and interoperability. Finally, Section X concludes with key insights on how orchestrated multi-agent systems can serve as a foundation for enterprise-scale AI ecosystems.

\section{Evolution of Agentic Systems}

Agentic systems are founded on the principle of autonomous entities that can perceive their environment, make decisions, and take actions to achieve specific goals. Defined by autonomy, reactivity, proactivity, and social ability, they extend beyond scripted automation to operate adaptively.  

Early deployments relied on single agents. These monolithic systems, with no coordination overhead, were well-suited to narrow tasks such as customer support chatbots that resolve FAQs, financial bots generating daily reports, or personal productivity assistants handling email and calendar management. Their reliability made them effective in bounded contexts, but they lacked scalability and adaptability for complex or dynamic environments. 

\begin{figure}
    \centering
    \includegraphics[width=1\linewidth]{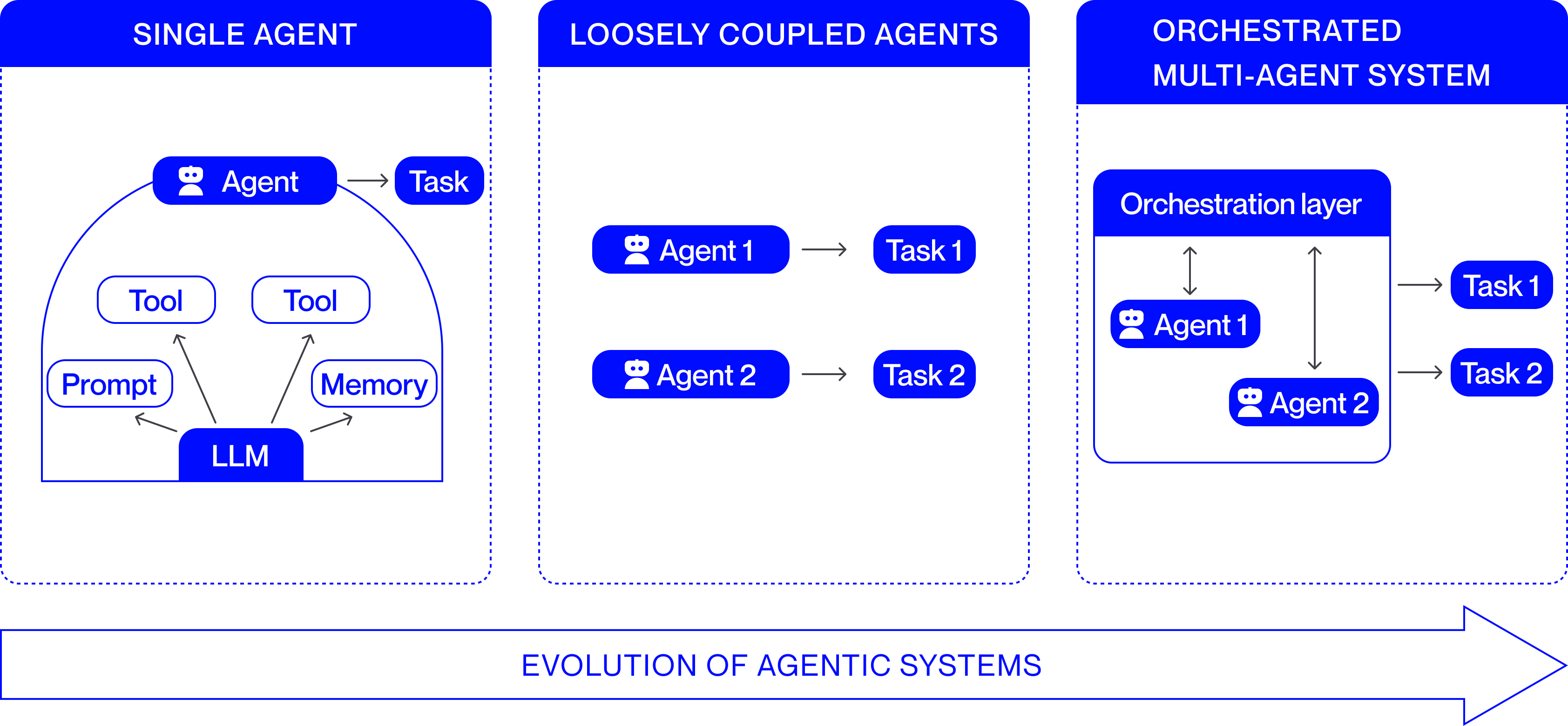}
    \caption{Evolution of Agentic Systems}
    \label{fig1}
\end{figure}

To address these limitations, research and practice shifted toward loosely coupled agentic systems. In such systems, multiple agents operate in parallel with relative independence, and minimal interaction. This architecture enables specialization and the emergence of collective behaviors that a single agent cannot achieve. Recent examples include scientific research assistants \cite{agentlab25} where literature-retrieval, reasoning, and validation agents collaborate to accelerate discovery; collaborative AI coding environments\cite{agentcoder23} where distinct agents write, review, and test code; news pipelines \cite{factcheck25} where aggregation, fact-checking, and synthesis are distributed across agents; and autonomous driving ecosystems ~\cite{koma24} where perception, navigation, and coordination agents cooperate loosely to ensure safe operation. 
Fig.~\ref{fig1} represents the discussed evolution of agentic systems.

\section{Architectural Composition of Multi-Agent Systems}

Building an orchestrated multi-agent system (MAS) involves more than simply connecting multiple autonomous agents. It requires designing specialized roles, establishing a coordination layer that governs their interactions, and defining clear communication protocols that allow agents to exchange information effectively.  

Specialized agents operate as autonomous components that execute well-defined tasks within the system, each contributing a distinct capability toward the collective objective. Their interactions are coordinated through an orchestration layer that determines execution order, manages dependencies, and aligns individual outputs into a coherent operational flow. The exchanges rely on communication protocols that standardize how information is represented and transferred across agents, ensuring semantic consistency and enabling the orchestration layer to maintain synchronized and interpretable system behavior. When these foundational components are applied in practice, they collectively enable complex, domain-specific workflows that require coordinated intelligence across multiple decision points.

The following sections examine these technical elements in greater depth, focusing on agent specialization, orchestration mechanisms, and communication strategies as the foundational components of multi-agent architectures. 

\section{Specialized Agents}

In a multi-agent system, specialized agents are autonomous components designed to perform narrowly scoped, role-specific tasks within the broader architecture. Each agent typically incorporates a large language model as its cognitive core, enabling it to perceive inputs, reason about them, and act within clearly defined operational boundaries. By assigning distinct roles such as retrieval, reasoning, validation, or monitoring, the system decomposes complex objectives into smaller, coordinated subtasks. This division of labor promotes modularity and collaboration, allowing agents to complement one another’s capabilities, reduce redundancy, and achieve outcomes that surpass those of a single, general-purpose agent. Through such specialization, a multi-agent system attains higher precision, scalability, and robustness while preserving clarity of function and accountability across its components. 

Below are the key categories of specialized agents:
\begin{itemize}
\item Worker Agents - Worker agents represent the most basic but essential type. They are responsible for carrying out well-defined tasks such as a Retrieval-augmented generation (RAG) pipeline. In practice, some worker agents are stateless, handling each request independently without retaining context, while others are stateful, tracking progress across multiple steps in a workflow. In large systems, worker agents often operate in parallel, with each specializing in a narrow sub-domain. In a financial underwriting workflow, individual worker agents may extract applicant data from loan documents, compute preliminary credit scores, or generate draft risk assessments that downstream agents validate and consolidate. These agents form the execution layer of the system, performing domain-specific computations that feed subsequent validation and oversight processes. 
\item Service Agents - Service agents provide shared operational capabilities that other agents depend on during workflow execution. They act as reusable utilities within the multi-agent ecosystem, performing tasks such as quality assurance, compliance enforcement, diagnostics, or automated recovery.

In the context of financial underwriting,  quality assurance agents can verify extracted customer data and cross-check it against compliance requirements. Diagnostic agents can inspect inconsistencies or missing data, trace the issues to the responsible module, and generate a structured error report that informs corrective actions. Healing agents can extend this capability by rerunning failed extractions or  resetting workflow states to restore normal operation. While healing agents focus on the active system state, upgrade scheduler agents manage version transitions of scoring components and related workflows, ensuring that  updates are deployed seamlessly without disrupting ongoing operations.

\item  Support Agents -  Complementing the service agents, support agents operate at a supervisory and analytical level. While service agents provide in-line operational utilities during workflow execution, support agents focus on meta-level oversight by monitoring system behavior, analyzing outcomes, and managing data flows that inform orchestration and optimization. Their function is to maintain the overall health, transparency, and adaptability of the system under varying operational conditions. In the financial underwriting scenario, monitoring agents track decision latency, detect risk model drift, and visualize overall portfolio health for both AI orchestrators and human supervisors. Analytics agents evaluate approval-rate patterns and compliance anomalies, while data agents refresh applicant datasets to maintain currency and accuracy.

\end{itemize}

\begin{figure}
    \centering
    \includegraphics[width=1\linewidth]{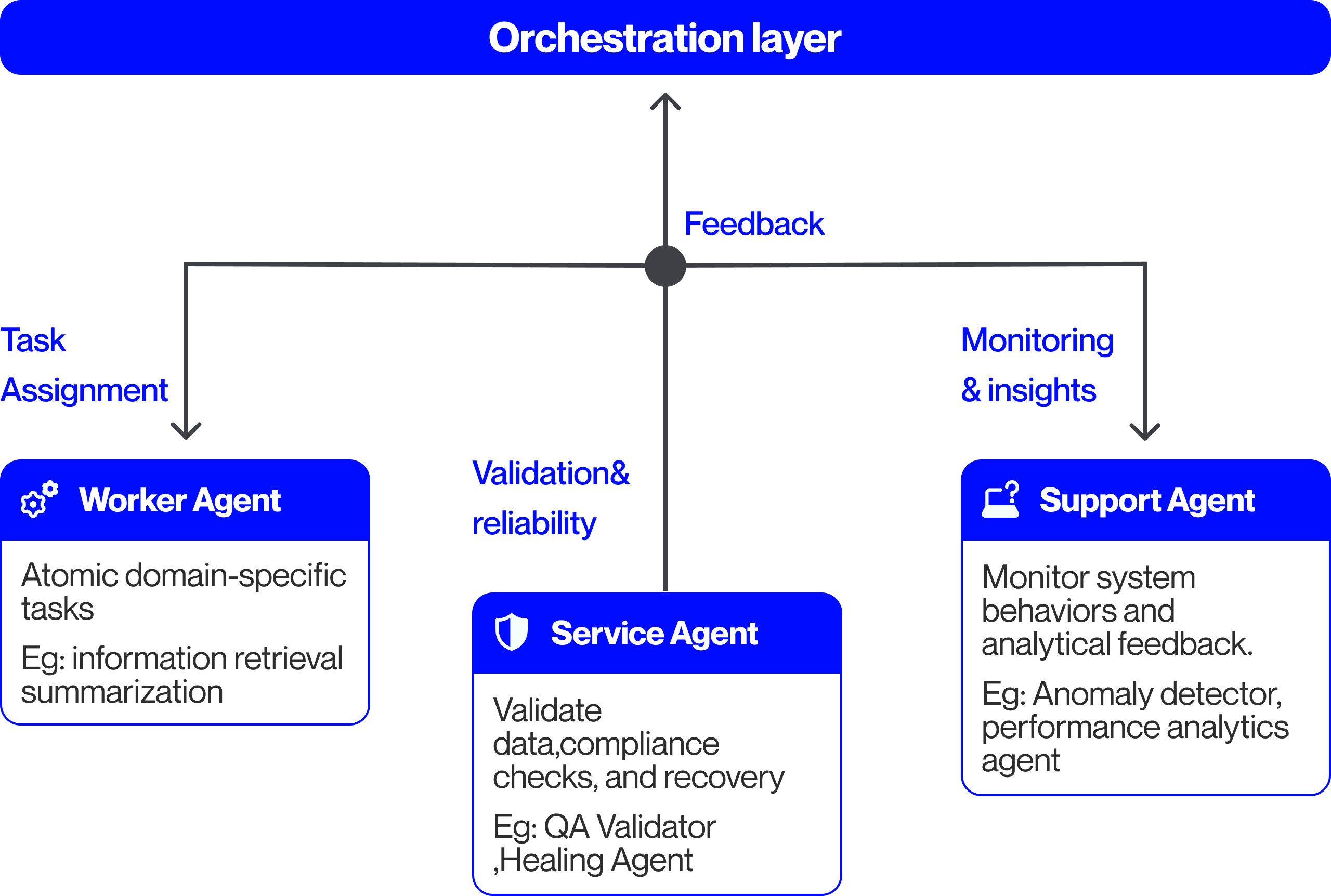}
    \caption{Specialized agents in a Multi-Agent System}
    \label{fig3}
\end{figure}

Fig.~\ref{fig3} summarizes the discussed categories of specialized agents in a multi-agent systems. Across these categories of specialized agents, recent research \cite{fu24} and enterprise applications demonstrate that coordinated role differentiation significantly enhances the reliability and scalability of multi-agent systems.

\section{Orchestration Layer for Coordinated Multi-Agent Operations}

The agent orchestration layer forms the control plane of a multi-agent system, transforming autonomous components into a coherent, goal-directed collective. Without orchestration, even highly capable agents risk duplication of effort, logical inconsistency, or unbounded autonomy that diverges from the system’s objectives. It interprets system-level objectives, decomposes them into actionable subtasks, coordinates their execution, and ensures that every output produced aligns with policy, context, and quality requirements. 

\subsection{Planning and Policy Management}

At the heart of the orchestration layer are the planning and policy units, which convert high-level objectives into a structured execution plan. 

Their purpose is twofold: the planning unit operates as a goal-decomposition engine that determines what tasks need to be done and in what order, while the policy unit embeds domain and governance constraints to define how these tasks are performed. In a financial underwriting workflow, the planning unit assigns tasks such as data extraction and credit scoring to the corresponding worker agents. The policy unit defines operational and governance constraints and coordinates with the control unit (described in the following subsection) to ensure service agents enforce them throughout execution  

Together, they translate abstract goals into a directed execution model that defines who performs which task, in what sequence, under what rules, and with what oversight. 

\subsection{Execution and Control Management}

Once tasks are planned and assigned, the orchestration layer operates as a distributed control system that transitions specialized agents through phases of initialization, execution, validation, and completion.  

Here, the execution unit ensures the smooth operation of the designated tasks performed by worker agents and manages telemetry data collection by support agents. The telemetry data is relayed to the control unit, which may delegate remediation or verification tasks to service agents to maintain operational stability. The control unit also manages concurrency and dependency across workflows, allowing parallel execution and synchronization at key checkpoints to preserve consistency. Additionally, it handles task prioritization and dynamic resource allocation to balance throughput, cost, and determinism across varying workloads. 

\subsection{State and Knowledge Management}

For the control unit to achieve synchronization and maintain continuity across workflows, the orchestration layer relies on the state and knowledge management component. This component functions both as a data bus and a knowledge repository.  

The state unit manages checkpoints, workflow progress, agent states, and activity logs. Support agents monitor state changes and performance anomalies, while service agents may be invoked to restore checkpoints from the state unit to preserve workflow integrity. The knowledge unit manages contextual and domain-specific information by connecting to external data sources and exposing them as a retrievable context. This ensures that worker agents and orchestration components operate with accurate and aligned information.  

This separation of operational state from knowledge state preserves modularity, contextual consistency, and system coherence. 

\subsection{Quality and Operations Management}

Using telemetry, state updates, and contextual data generated by other orchestration units, the quality and operations management component evaluates system performance, validates outcomes, and ensures that orchestrated activities remain compliant and continually optimized. While the control unit focuses on execution stability and the policy unit defines and enforces operational constraints during execution, this component governs verification and optimization of results after execution across the orchestration layer.  

The component validates aggregated outputs against defined schemas before integrating them into the shared state, preventing invalid data from propagating through workflows. When inconsistencies or violations are detected, it updates the state accordingly and may invoke service agents to perform diagnostic or remediation actions, ensuring sustained integrity and compliance. 

The subsystem also monitors metrics such as latency, throughput, and success rate, using anomaly detection to identify deviations and trigger preemptive interventions. It also supports controlled deployment, testing, and sandboxing new components to maintain stability as agents evolve. Together, these mechanisms ensure resilient, auditable, and continuously improving multi-agent operations. 

\subsection{Closing Discussion}

The orchestration model is exemplified in a financial institution’s credit-risk and fraud detection workflow, where specialized agents are coordinated to ensure consistency and compliance.  Incoming loan applications are decomposed by the planning unit into subtasks such as data extraction, risk assessment, compliance review, and fraud screening, while the policy unit embeds governance constraints, including lending regulations and institutional risk thresholds. The execution and control component manages concurrent task execution, collects telemetry, and invokes service agents for recovery when needed. The state and knowledge management component maintains applicant states, historical records, and regulatory references for contextual continuity, while the quality and operations component validates results against policy criteria and applies performance insights to optimize future workflows. 

 Collectively, these mechanisms show that reliability in multi-agent systems arises not only from intelligent agents but from the orchestration layer that governs planning, execution, and validation, enabling scalable and policy-compliant performance.  

\section{Communication Protocols in Orchestrated Systems}

Building on the orchestration framework, the agent communication layer operationalizes coordination by enabling agents and external tools to exchange information, control signals, and shared context. While orchestration defines who acts and when, communication ensures those actions remain synchronized and interpretable. Traditional protocols relied on static request–response exchanges and lacked mechanisms for context sharing or policy enforcement. To address these limitations, two emerging standards—the Model Context Protocol (MCP) and Agent-to-Agent (A2A) protocol—establish structured, interoperable communication for tool interaction and peer specialized agentic collaboration. Later subsections explore these protocols in more detail.  

\subsection{Model Context Protocol}

\begin{figure}
    \centering
    \includegraphics[width=1\linewidth]{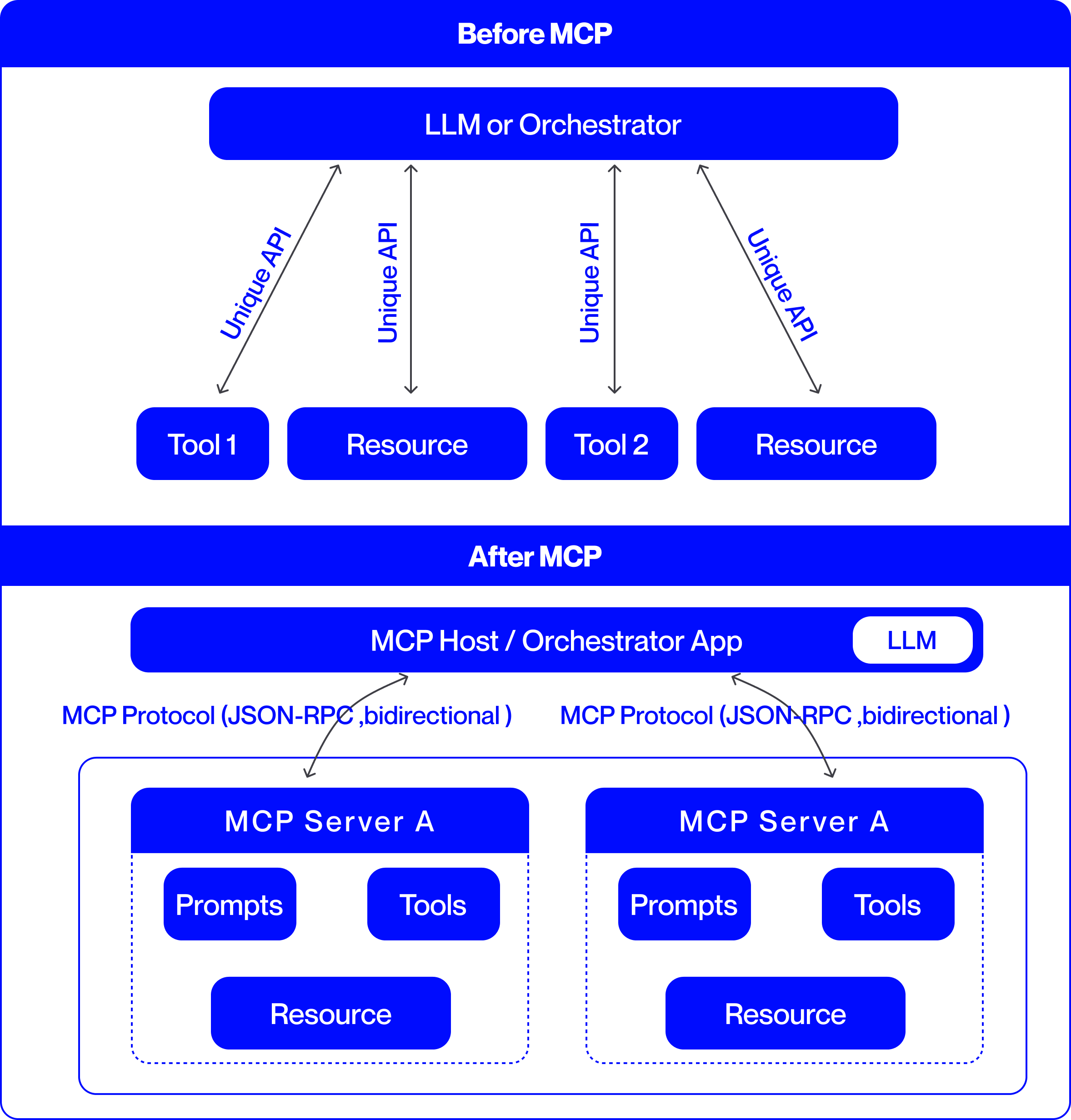}
    \caption{Comparison of agent communication without MCP and with MCP}
    \label{fig5}
\end{figure}

The Model Context Protocol (MCP)\cite{mcpdocs25} provides the standardized communication interface that operationalizes the previously described orchestration flow between agents and external systems like tools, data services, and contextual repositories. As illustrated in Fig.~\ref{fig5}, MCP mediates every external invocation through a defined interface that enforces schema consistency, access control, and auditability. This allows the execution unit to dispatch tasks confidently, ensuring that each agent’s external interactions conform to orchestration policies and operational rules.

\begin{figure}
    \centering
    \includegraphics[width=1\linewidth]{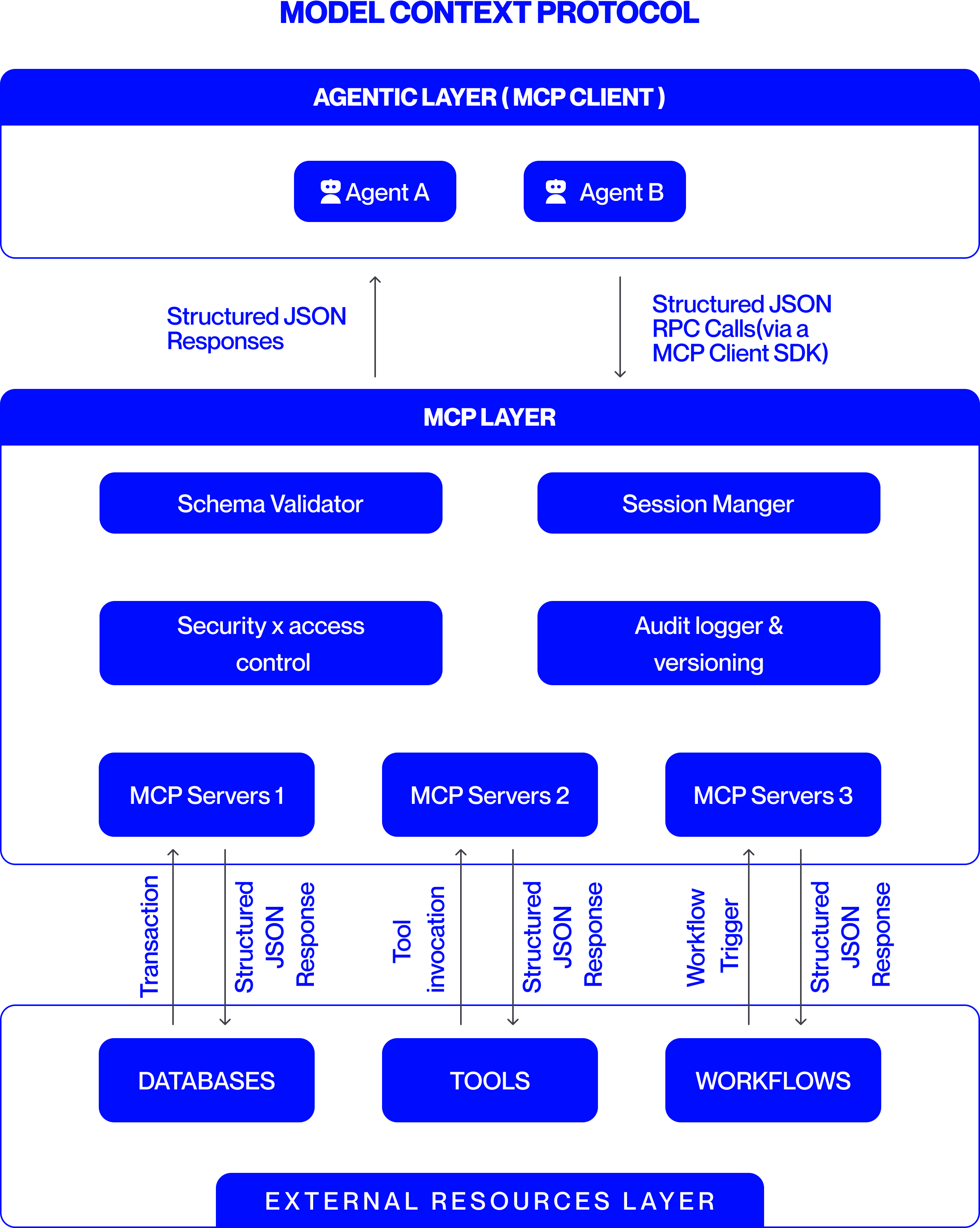}
    \caption{Integration of the MCP within the orchestration architecture}
    \label{fig6}
\end{figure}

Within the orchestration architecture, MCP follows a client–server design in which agents or orchestrators act as clients that request external capabilities such as tools, resources, or prompts, while connected systems expose these as standardized, callable services. Through MCP, the planning and control units translate defined tasks into executable tool calls and govern how and when agents access these resources in compliance with policy constraints. MCP’s session management supports both stateless and stateful exchanges, allowing context continuity across multi-step workflows. These exchanges are logged and synchronized with orchestration state, enabling the state and knowledge management component to maintain consistency and allowing the quality and operations component to verify compliance and alignment with expected results. 

As shown in Fig.~\ref{fig6}, MCP functions as the operational bridge between high-level orchestration plans and low-level tool execution. It converts planned objectives into structured, policy-aligned invocations and feeds execution data back into orchestration memory and quality loops. Extensions such as ScaleMCP\cite{scalemcp25} dynamically synchronize tool inventories across agents, while AgentMaster\cite{agentmaster25} integrates MCP with inter-agent communication frameworks such as A2A to support multimodal collaboration and information retrieval within orchestrated multi-agent systems. 

\subsection{Agent-to-Agent Protocol}

\begin{figure}
    \centering
    \includegraphics[width=1\linewidth]{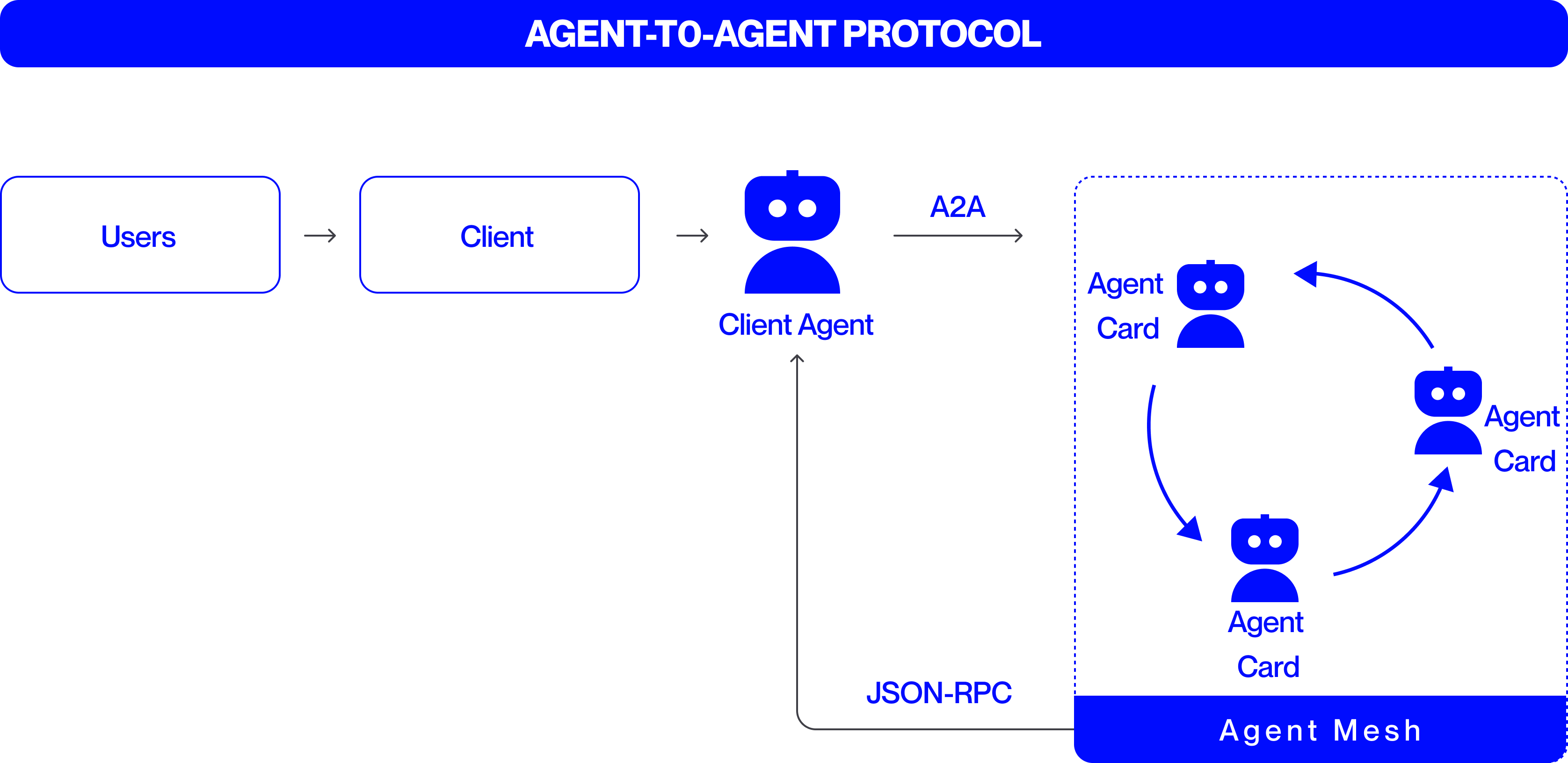}
    \caption{Agent-to-Agent Protocol}
    \label{fig7}
\end{figure}

While MCP governs how agents interact with tools and data, the Agent-to-Agent (A2A) protocol \cite{a2aguide25} defines standardized communication amongst specialized agents themselves. It supports negotiation, delegation, and coordination across distributed ecosystems while maintaining interoperability, traceability, and security \cite{a2aprotocol25}. Together, MCP and A2A form the dual foundation of agent communication—MCP for tool access and A2A for peer collaboration (Fig.~\ref{fig7}).

Through A2A, worker agents can delegate subtasks or share intermediate results, service agents can communicate diagnostic information or recovery status, and support agents can broadcast telemetry or performance insights that inform collective progress. This peer-level exchange ensures that task dependencies are dynamically managed and that agents can resolve interdependencies without requiring centralized intervention. The control unit supervises these interactions to ensure policy alignment and to maintain synchronization with the broader orchestration plan, while communication records are captured within the state and knowledge management component for traceability and recovery. 

A2A employs a peer communication model either direct or mediated through the orchestrator, enabling reliable, authenticated message exchange.  Each message  carries structured metadata and standardized payloads, ensuring consistency across heterogeneous implementation.  Robust security controls, including cryptographic signing and role-based routing, guarantee message integrity and policy compliance. 

Although peer-oriented, A2A remains supervised by the orchestration layer, which validates and synchronizes exchanges to maintain coherence with global workflows. Emerging research explores scalable and hybrid architectures that combine A2A and MCP for multimodal, adaptive coordination in enterprise-grade agentic systems.

\section{Safety, Governance and Observability}

Ensuring the reliability of multi-agent systems depends on safeguards embedded within orchestration and communication mechanisms. Within the orchestration layer, the control and quality and operations management units enforce safety and governance through validation, monitoring, and recovery mechanisms that maintain compliance and operational integrity. Similarly, MCP and A2A protocols embed protective measures such as schema validation, authenticated exchanges, and access control to ensure secure and interpretable communication. Core guardrails mitigate hallucinations and enforce consistency checks to prevent agents from producing unsafe or conflicting outputs. These protections are reinforced by internal audits, event logging, and least-privilege policies that promote transparency, accountability, and traceability across the system. Privacy constraints restrict agents to sharing only task-relevant information. Continuous monitoring, carried out through support agents and the quality and operations management unit, tracks latency, throughput, and correctness to evaluate performance and detect drift. Together, these practices transform multi-agent systems from experimental collectives into dependable, auditable, and continually improving infrastructures that balance autonomy with control.

\begin{figure}
    \centering
    \includegraphics[width=1\linewidth]{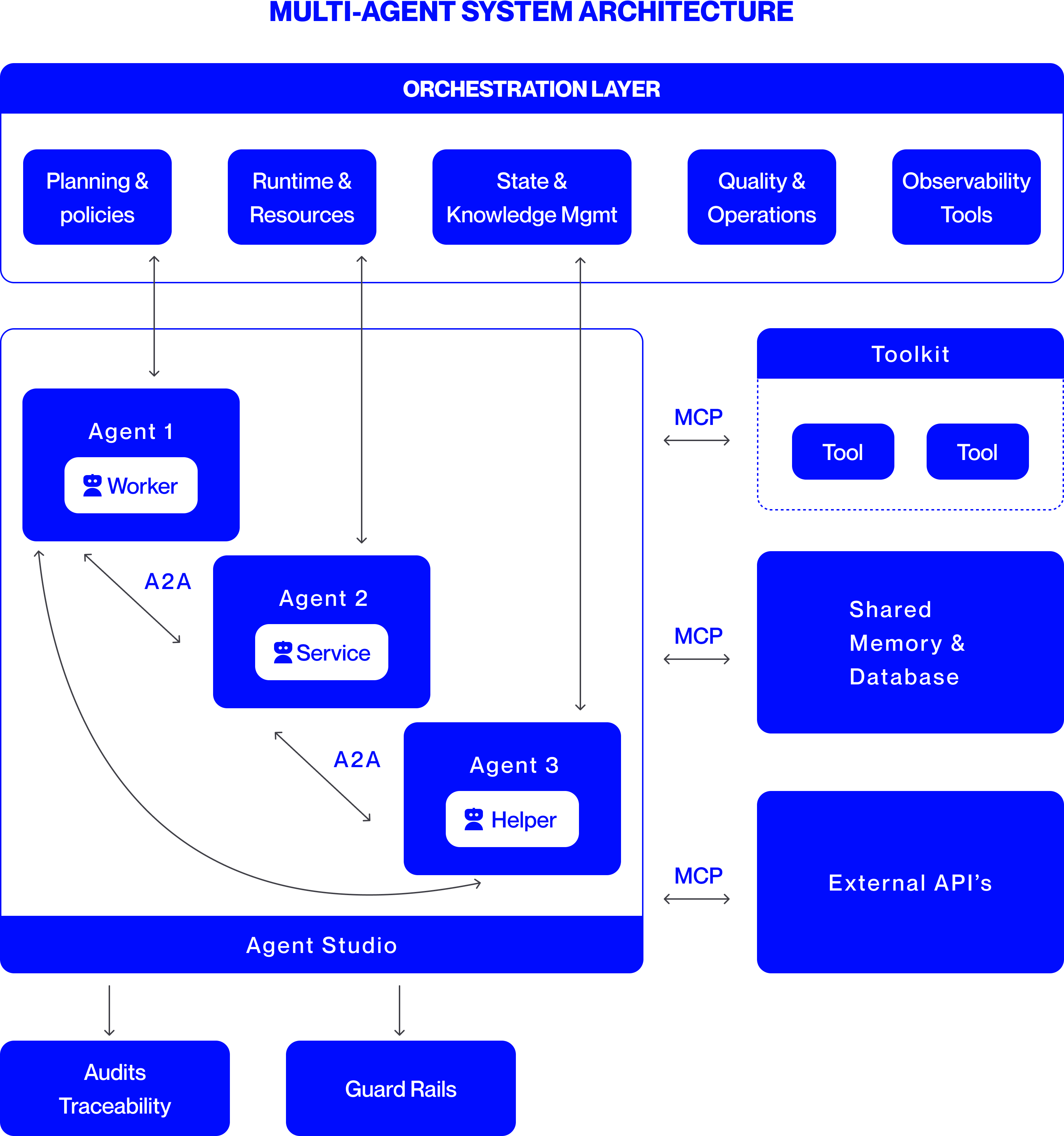}
    \caption{Orchestrated Multi-Agent System Architecture}
    \label{orchestratedMAS}
\end{figure}

To summarize, the overall architecture of an orchestrated multi-agent system is shown in Fig. ~\ref{orchestratedMAS}. The architecture integrates all core components that enable coordination, communication, and governance across distributed agents. At its foundation are specialized agent types, interacting through standardized protocols such as MCP for tool and data access and A2A for inter-agent collaboration. The orchestration layer oversees planning, execution, and quality control, while observability and governance modules ensure reliability, compliance, and transparency. Together, these elements form a cohesive, scalable framework that operationalizes autonomous intelligence under structured orchestration.

\section{Case Studies}

\subsection{Banking, Financial Services and Insurance}

Multi-agent AI systems are revolutionizing Banking, Financial Services, and Insurance (BFSI) industry by delivering dramatic efficiency gains and ROI. Insurers are deploying networks of specialized AI agents to automate the labor-intensive underwriting process. For example, autonomous agents studied in \cite{multimodal25} now parse insurance applications and supporting documents with over 95\% accuracy, enabling much faster policy issuance. In another use-case explored in \cite{multimodal25}, a mortgage lender integrated Document AI and Decision AI agents to handle loan paperwork, achieving a 20× faster approval process while cutting processing costs by 80\%.  Similarly,\cite{scirp25} presents a multi-agent automation framework for property claims underwriting, where specialized agents collaborate to evaluate claim documents, assess damage estimates, and validate policy conditions. Such multi-agent systems clearly outperform both manual processes and single-agent automation, delivering a strong value proposition in BFSI. 

\subsection{Software Engineering and IT Modernization}

Multi-agent systems are also proving their worth in software engineering. A large bank recently applied an agentic AI “digital factory” \cite{mckinsey25} to modernize its legacy core software, which comprised hundreds of applications. Different agents took on specialized coding tasks: one agent automatically documented existing legacy code, another generated new code modules, others reviewed code written by their peers, and additional agents integrated and tested these modules. This multi-agent architecture allows parallel execution and continuous code quality checks, reducing the coordination burdens that slowed the purely human teams. In practice, this approach led to over a 50\% reduction in development time and effort for early-adopter teams at the bank.

\subsection{Cross Industry Adoption}

The success of multi-agent AI is spurring widespread adoption across industries. In customer service, companies are reimagining call centers with agentic AI: instead of just assisting human representatives, agents autonomously handle routine inquiries and issues. Studies suggest that up to 80\% of common support incidents could be resolved by AI agents without human intervention, cutting resolution times by 60–90\% in fully agent-driven workflows. Meanwhile, sectors like healthcare are exploring multi-agent setups where one agent analyzes patient symptoms or medical literature and another suggests treatment plans, all under a doctor’s supervision. From finance and insurance to software development, legal research, and healthcare, multi-agent systems are being rapidly embraced as organizations recognize their measurable performance advantages over manual or single-agent approaches.

\section{Challenges, Risks and Future Research}

As multi-agent systems scale, key challenges emerge around efficiency, cost, and governance. Coordination among numerous agents can create communication overhead, message congestion, and performance bottlenecks unless workflows are carefully managed. The cost of adoption remains significant, enterprises must invest in orchestration software, skilled engineering teams, and continuous monitoring infrastructure to ensure reliability and compliance. Governance presents another difficulty, as decentralized autonomy complicates oversight and accountability. Risks inherited from large language models, such as hallucination, bias, and data leakage, are magnified when agents interact, raising safety, ethical, and privacy concerns that demand rigorous evaluation and control frameworks.

Future research focuses on making orchestration smarter, safer, and more adaptive. Hybrid and federated designs aim to balance centralized control with decentralized flexibility, while semantic orchestration seeks to match tasks dynamically with the most capable agents. Advances in federated learning and cross-domain collaboration promise secure knowledge sharing without exposing raw data. Standardized benchmarks, simulation testbeds, and open-source orchestration frameworks will further enable transparent performance comparison and lower entry barriers. Together, these directions move multi-agent systems toward scalable, accountable, and trustworthy deployment across enterprise and societal domains.

\section{Conclusion}

Agentic systems have evolved from single agents that perform narrow tasks, to loosely coupled multi-agent setups, and now to orchestrated collectives where coordination ensures consistency, scale, and reliability. Recent advances show that orchestrated systems are not only viable but already delivering value in real deployments, from BFSI claims processing and fraud detection to healthcare diagnostics and software engineering. Benchmarks and case studies demonstrate measurable gains in productivity, error reduction, and scalability compared with manual or single-agent approaches. 

Looking forward, enterprises are moving toward dynamic ecosystems where agents can form, dissolve, and reorganize in response to tasks, much like human teams. To realize this vision, the community must invest in open protocols for interoperability, standardized benchmarks, and shared research infrastructure. With these foundations, orchestrated multi-agent systems can mature into a reliable and adaptable backbone for enterprise intelligence at scale.

\end{document}